\documentclass[11pt,twoside]{article}
\usepackage{asp2010}

\resetcounters

\begin{document}

\title{Young Stars in the Time Domain: A CS16 Splinter Summary}
\author{Kevin R. Covey$^{1,2}$, Peter Plavchan$^3$, Fabienne Bastien$^4$, Ettore Flaccomio$^5$, Kevin Flaherty$^6$, Stephen Marsden$^7$, Maria Morales-Calder\'on$^8$, James Muzerolle$^9$, Neal J. Turner$^{10}$ 
\affil{$^1$Hubble Fellow; Cornell University, Department of Astronomy, 226 Space Sciences Building, Ithaca, NY 14853, USA.}
\affil{$^2$Visiting Researcher, Department of Astronomy, Boston University, 725 Commonwealth Ave, Boston, MA 02215, USA.}
\affil{$^3$NASA Exoplanet Science Institute, California Institute of Technology, MC 100-22, 770 S. Wilson Avenue, Pasadena, CA 91125, USA.}
\affil{$^4$Department of Physics and Astronomy, Vanderbilt University, Nashville, TN 37235}
\affil{$^5$INAF - Osservatorio Astronomico di Palermo, Piazza del Parlamento 1, 90134 Palermo, Italy.}
\affil{$^6$Steward Observatory, University of Arizona, Tucson, AZ 85721, USA.}
\affil{$^7$Center for Astronomy, School of Engineering and Physical Sciences, James Cook University, Townsville, 4811, Australia}
\affil{$^8$Spitzer Science Center, California Institute of Technology, Pasadena, CA 91125, USA.}
\affil{$^9$ Space Telescope Science Institute, 3700 San Martin Dr., Baltimore, MD 21218, USA.}
\affil{$^{10}$Jet Propulsion Laboratory, California Institute of Technology, 4800 Oak Grove Drive, Pasadena, CA 91109, USA.}
}

\begin{abstract}
Variability is a defining characteristic of young stellar systems, and optical variability has been heavily studied to select and characterize the photospheric properties of young stars. In recent years, multi-epoch observations sampling a wider range of wavelengths and time-scales have revealed a wealth of time-variable phenomena at work during the star formation process. This splinter session was convened to summarize recent progress in providing improved coverage and understanding of time-variable processes in young stars and circumstellar disks.  We begin by summarizing results from several multi-epoch {\it Spitzer} campaigns, which have demonstrated that many young stellar objects evidence significant mid-IR variability.  While some of these variations can be attributed to processes in the stellar photosphere, others appear to trace short time-scale changes in the circumstellar disk which can be successfully modeled with axisymmetric or non-axisymmetric structures.   We also review recent studies probing variability at shorter wavelengths that provide evidence for high frequency pulsations associated with accretion outbursts, correlated optical/X-ray variability in Classical T Tauri stars, and magnetic reversals in young solar analogs. 
\end{abstract}

\section{Introduction}

The formation and early evolution of stars is inherently a time-domain process: the conversion of a dense molecular core into a zero-age main sequence star requires nearly every pertinent physical parameter (radius, density, temperature, v$_{rot}$, etc.) to change by multiple orders of magnitude over the relatively brief timescale of 10s of Myrs.  These time-scales still dwarf that of a human lifetime\footnote{or, perhaps more relevantly, the timescale of a PhD thesis}, and one might {\it a priori} conclude that star formation is no more amenable to time-domain study than any other aspect of a stellar astrophysics.  

Photometric variability has nonetheless been recognized for decades as a common trait of many young stars \citep[e.g., ][]{Joy1945}.  Historically, this variability has been best explored via optical photometry, most often sampling time-scales of days to (a few) years.  These studies have provided detailed descriptions of the spot properties of optically revealed young stars \citep{Vrba1988}, a comprehensive inventory of stellar rotation in young clusters \citep[see first the review by][]{Herbst2007, Bouvier1995, Stassun1999, Rebull2004, Cieza2007, Irwin2008}, and a quantitative statistical portrait of accretion and extinction-induced variability \citep{Grankin2007}.  These monitoring programs have also identified a number of rare, astrophysically valuable systems: pre-main sequence eclipsing binaries \citep[e.g., ][]{Cargile2008}, disk occulting systems \citep[ie, KH-15D:][]{Hamilton2001}, and stars undergoing massive accretion events \citep[i.e., FU Ori, EX Lup, or V1647-like variables:][]{Herbig1977, Herbig1989, Hartmann1996, Reipurth2004, Lorenzetti2007}.

Significant advances in observational time domain astronomy are uncovering new phenomena and systems for study.  This expansion is due to many factors: a steady improvement in the size and sensitivity of optical arrays, enhancing the coverage and cadence possible for observations of optically revealed clusters; even greater advances in the capabilities of near-infrared arrays; and access to precise multi-epoch mid-infrared (mid-IR) photometry and spectroscopy from the {\it Spitzer Space Telescope}.  These new observational capabilities have allowed variability studies to cover a greater number of targets at higher cadences, and characterize the variability properties of more deeply embedded, and presumably less evolutionarily advanced, sources.  The signature of these advances is evident in our increased sensitivity to variability at the lowest masses within nearby star-forming regions \citep{Cody2010} as well as the increasing frequency with which we identify formerly rare young variables, including new eclipsing binaries \citep{Hebb2010}, new disk occulting systems \citep[e.g., WL4:][]{Plavchan2008}, pulsating protostars \citep{Morales-Calderon2009} and large-amplitude outbursts \citep{Covey2010,Miller2010,Kospal2010,Garatti2010}.  These observational advances have been accompanied by similar progress on the theoretical front, with increasingly detailed models of the physical processes underlying the observed variations \citep[e.g., accretion variations, disk processes, etc.;][]{Vorobyov2010, Zhu2010, Baraffe2010}

This splinter session was convened to review recent progress in characterizing, analyzing, and understanding variability in the youngest stars.  The session incorporated presentations covering physical processes that develop over a range of time-scales, with observational signatures spanning a wide range of wavelengths.  Following these presentations, the audience participated in a broader discussion of these new results and highlighted areas of particular promise for future observational or theoretical work.  We provide a brief summary of each presentation below, and conclude with a recap of the open questions highlighted in the audience discussion.

\section{Mid-infrared (Mid-IR) variability}

\subsection{Maria Morales-Calder\'on: A global view of Mid-IR variability from the YSOVAR Orion survey}

\begin{figure}[!ht]
\plotone{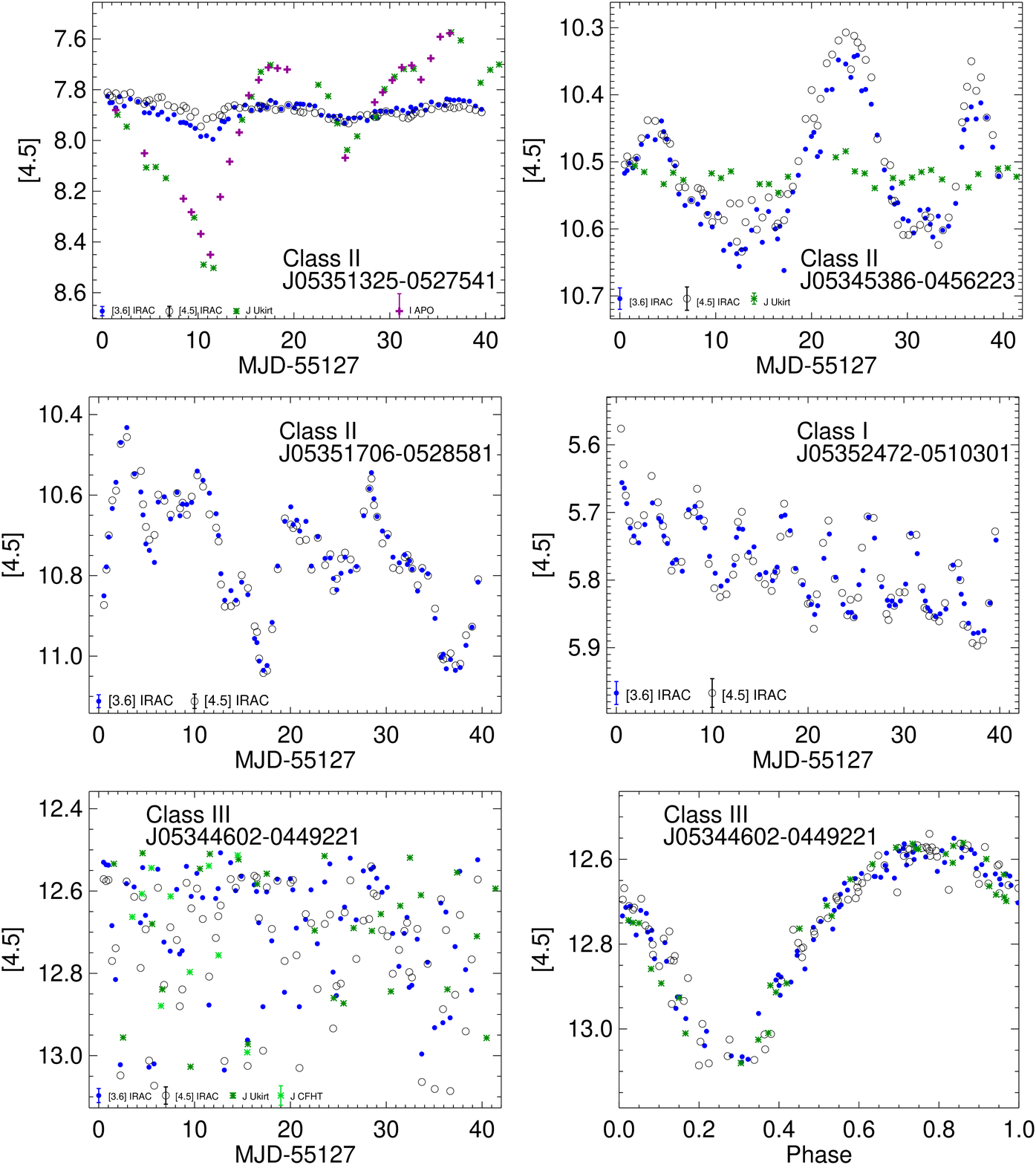}
\caption{A sample of the range of variability types captured by the YSOVAR program.  Most variations are relatively colorless (e.g., the bottom four panels), but some strong color variations are observed. Both red-dominant and blue-dominant color variations are detected (see top left and right panels, respectively), suggesting that different physical mechanisms (extinction, variable disk temperature \& geometry, etc.) may be required to explain the full range of color-dependent variability.  Aperiodic variability (which we typically interpret as disk/accretion driven) is the dominant mode of variability seen in the disked/embedded sources (e.g., middle-left panel), but we do see examples of periodic variability even in the most heavily embedded sources (e.g., middle-right panel).  Unphased and phased light curves are shown for a particularly short period WTTs system ($\sim$0.27 d) in the bottom right and left panels, respectively.  While we typically interpret periodic WTTs as exhibiting spot-induced variability, this system's short period and asymmetric light curve suggest it may be a close, tidally distorted binary.}
\label{YSOVAR_LCs}
\end{figure}

\begin{figure}[!ht]
\plotfiddle{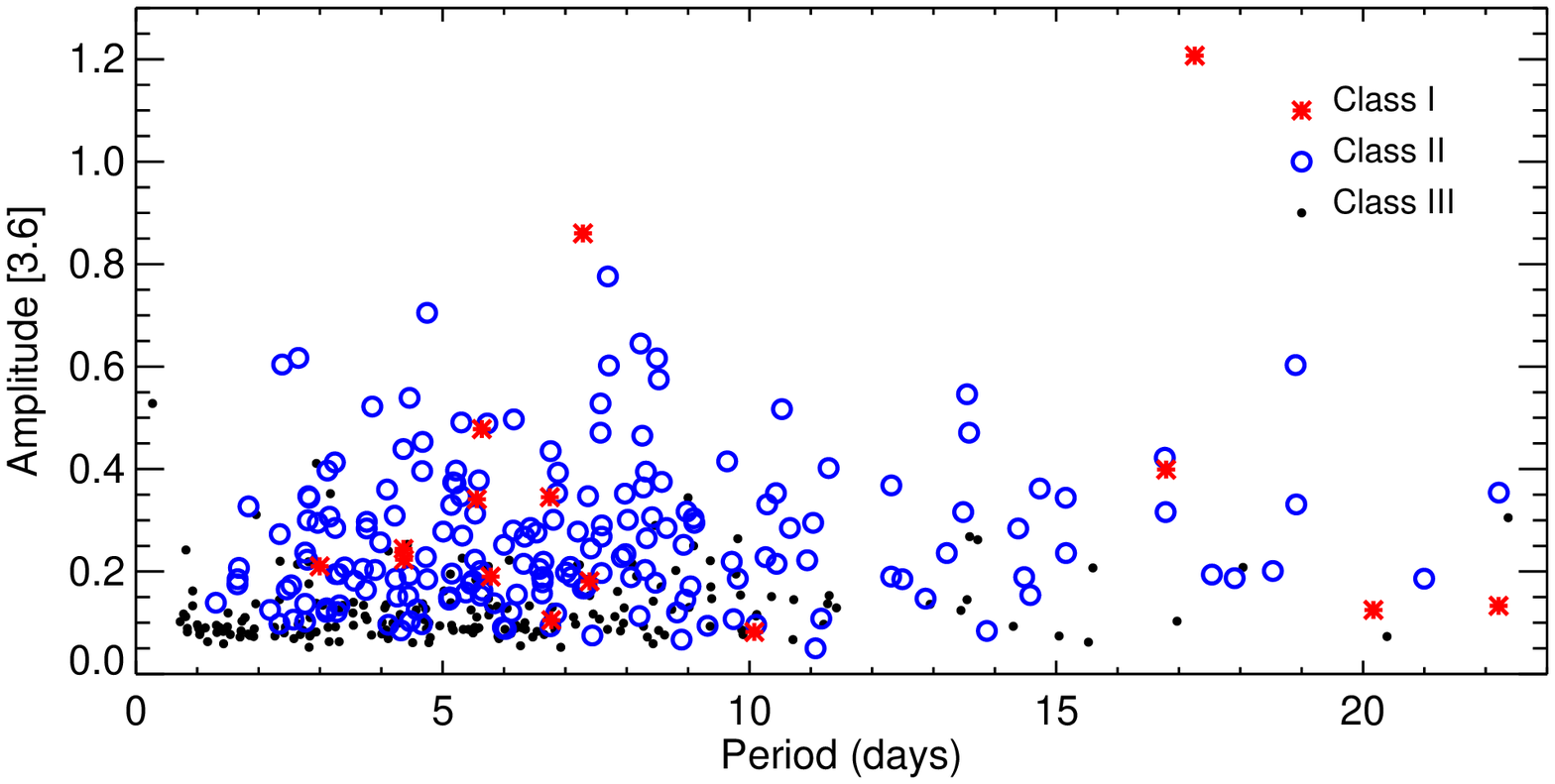}{2in}{0}{50}{50}{-160}{-20}
%\scale{0.7}
%\plotone{AmpvsPeriod.ps}
\caption{3.6 $\mu$m amplitude vs. period for YSOs detected as periodic variables in YSOVAR's mid-IR monitoring of Orion.  Both period and amplitude appear correlated with SED class: while periodic variability is most commonly detected for WTTs/Class III objects, those disked/embedded Class I-II YSOs that do demonstrate periodic variability have generally longer periods and larger amplitudes.}
\label{YSOVAR_periods}
\end{figure}

The YSOVAR Orion program provides the first large-scale survey of the photometric variability of Young Stellar Objects (YSOs) in the mid-IR. In Fall 2009, \textit{Spitzer}/IRAC observed a 0.9 sq. deg. area centered on the Trapezium cluster twice a day for 40 consecutive days, producing 3.6 and 4.5 $\mu$m high fidelity light curves ($\sim$3\% typical photometric uncertainties) for over 2000 disked and diskless YSOs in Orion (as diagnosed with precursor 3.6-8.0 $\mu$m IRAC photometry). For brevity, we refer to the diskless stars as Weak T Tauri stars (WTTs); we refer to stars with disks, or which are otherwise heavily embedded in their natal cloud, as YSOs.  For many of the stars, we also obtained complementary time-series photometry at optical (I$_c$) and/or near-infrared (JK$_s$) wavelengths. We find that 65\% of the disked YSOs and 30\% of the diskless WTTs are variable. The WTT variations are mostly spot-like, while the disked/embedded YSOs display a wider range of variability types: see Figure \ref{YSOVAR_LCs} for examples of the different types of variability evident in the YSOVAR light curves.  Consistent with their greater propensity for spot-like variability, 65\% of the variable WTTs are periodic, while periods are detected for only 16\% of the YSOs. Those disked/embedded YSOs that are detected to be periodic, however, typically evidence larger amplitudes, as well as longer periods (see Figure \ref{YSOVAR_periods}).  

YSOVAR monitoring has also identified several rare pre-main sequence variables.  Among these are five new candidate PMS eclipsing binaries, including one that is fainter than all previously known ONC PMS eclipsing binaries.  One of the most surprising and interesting classes of variables we find are characterized by short duration flux dips \citep[less than one to a few days; ie, AA Tau analogs:][]{Bouvier2003}. We interpret these events as the star being extincted by either clouds of relatively higher opacity in the disk atmosphere, or geometric disk warps of relatively higher latitude which pass through the line of sight to the star. This hypothesis is consistent with both the duration and the wavelength dependence of the events (i.e., larger amplitudes at bluer wavelengths).

\subsection{James Muzerolle: Spitzer Mid-Infrared Spectroscopic Variability in IC348}

We initiated a multi-epoch study of the young cluster IC 348 with all instruments on board \textit{Spitzer}.  This region is an appealing target given its reasonably close distance and wide range of YSO evolutionary states, including a significant number of highly evolved and transitional disks.   With prior GTO/legacy observations, our dataset includes a total of 10 maps of the entire cluster with MIPS, 7 maps with IRAC, and 4 to 7 epochs of IRS spectroscopy for 14 individual cluster members \citep[InfraRed Spectrograph;][]{Houck2004}.  The measurements sample a range of cadences from days to years over the period 2004 - 2009.  We find ubiquitous variability at all wavelengths: over half of the Class 0/I objects at 3.6 - 24 $\mu$m, nearly 70\% of Class II objects at 3.6 - 8 $\mu$m, and 40\% of Class II objects at 24 $\mu$m.  Variability time-scales range from days to years at all wavelengths.  We see apparent wavelength dependences with two general flavors: correlated variability over the 3.6 - 24 $\mu$m range, with or without color changes (most common in Class 0/I sources), and anti-correlated behavior with a pivot point typically around 6 - 10 microns (most common in Class II sources and transitional disks).  From the IRS spectra, almost all of the variations are related to the continuum; we see measurable changes in the silicate features in only a few objects.

In most cases, the variations seen at longer wavelengths occur on time-scales too short to be representative of in-situ structural changes.  For the Class II objects with anti-correlated wavelength behavior, the flux changes are likely connected to the innermost disk regions (puffed inner rim associated with dust sublimation, or dynamical truncation from close companions).  Perturbations can cause a change in the inner disk height, which leads to direct changes in the short wavelength emission, and can cause shadowing of cooler material farther out in the disk, which leads to changes in the longer wavelength emission.  The origin of the inner disk perturbations is not clear, but may be a result of changes in the accretion luminosity emitted by stellar accretion shocks, or inner disk warping created by tilted stellar magnetic fields or gravitational interactions with embedded companions.

\subsection{Kevin Flaherty: LRLL 31, a case study of Mid-IR variability in transition disks}

In addition to providing a robust characterization of mid-IR photometric variability, the {\it Spitzer Space Telescope} has diagnosed mid-IR variability in specific young stars via multi-epoch spectroscopy with IRS.   These observations identified the detailed wavelength dependence of mid-IR flux variations, pointing the way to new and unexpected results.  One of the most interesting cases is that of LRLL 31, a G6 T Tauri star in IC 348: IRS spectra separated by one week revealed that the 5-8\micron\ flux decreased to nearly photospheric levels while the 8-40\micron\ flux increased by 60\% \citep{Muzerolle2009}.  What makes this star especially interesting is that it shows a deficit of flux around 10 \micron\ compared to a traditional T Tauri star, indicative of a removal of small dust grains from the inner disk. This transition disk represents an evolutionary stage where the dust and gas are being removed from the circumstellar disk, and the fluctuations may be related to the process that is clearing the disk. Previous theories for explaining variability in young stellar objects (rotation of hot/cold spots across the stellar surface, changes in the accretion rate, extinction events) do not explain the wavelength dependence or the strength of the variability observed in LLRL 31. Fluctuations on daily to weekly time-scales are surprising, since thermal equilibrium arguments would indicate that mid-IR wavelengths probe regions of the disk from 0.5 to a few AU, where the dynamical timescale is closer to years. {\it In situ} changes in this region of the disk should not be possible on these time-scales, so LLRL 31's mid-IR variability must be rooted in the dynamics of the inner disk.

Motivated by these observations of LLRL 31, \citet{Flaherty2010} developed a model that explains the observed variability as arising from a warp in LLRL 31's disk, as opposed to the purely axisymmetric disks that had been previously assumed. Varying the height of a warp at the inner edge of the disk was able to explain the observed strength and wavelength dependence of the variability. As the warp grows it is more directly illuminated by the central star and heats up, emitting more flux at shorter wavelengths. The larger warp also shadows more of the outer disk, reducing the long-wavelength flux that originates far from the star. This model assumes the variability arises from structural changes in the disk warp: the precession of such a structure is unable to reproduce the variability.

While this model was successful at explaining much of LLRL 31's variability at $\lambda<10 \micron$, it was still limited by the fact that the 5-40 \micron\ IRS spectra only partially traces emission from the inner disk. Observations at shorter wavelengths, produced by emission from the hottest dust, are needed to better understand the properties of the material responsible for the mid-infrared variations. Further monitoring at 3.6 and 4.5 \micron\ with Spitzer, and ground-based 0.8-5 \micron\ spectra have revealed that the temperature of LLRL 31's inner disk remained constant while the emitting area changed by a factor of 10. Moreover, in 2009 there appeared to be a correlation between the accretion rate, as measured by the Pa$\beta$ line, and the infrared flux, measured by the infrared photometry \citetext{Flaherty et al. submitted}. This strong correlation suggests that the process that disturbs the dust must also disrupt the accretion flow. The most promising explanations are either a companion beyond 0.3 AU on an orbit misaligned with the disk \citep[e.g. ][]{Fragner2009} or a dynamic interface between the stellar magnetic field and the disk \citep[e.g.,][]{Goodson1999, Bouvier2007}. Contemporaneous X-ray and Spitzer observations will be able to trace some of the interaction between the magnetic field and the disk, further probing this process. The wavelength dependence of the variability seen in LRLL 31 appears to be common among transition disks \citetext{Espaillat et al. submitted} and further observations of both the gas and dust in these objects could determine if the physical process occurring in LRLL 31 is common in these other objects.

\subsection{Neal Turner: Models of axisymmetric disk disturbances from thermal waves and turbulent eddies}    

T Tauri stars' diverse mid-IR variability likely arises from several
different processes.  Variations that correlate poorly with the
accretion luminosity are most simply explained by changes in the way
the starlight falls across the disk surface.  Kevin Flaherty presents above
some ways to produce non-axisymmetric disk disturbances.  Among
processes leading to approximately-axisymmetric disturbances, two
explored here are thermal waves and turbulent eddies.

Thermal waves are driven by the stellar illumination.  The front of
each wave tilts into the starlight, receiving extra heating and
expanding, while the back of each wave receives less heating and cools
and contracts.  The wave propagates inward, dying out when it enters
the shadow of the disk's inner rim.  The wavelength is about equal to
the length of the shadow cast by each peak.  Thermal waves operate at
mass accretion rates below about $10^{-7}$ \.{M}$_{\odot}$/yr.
For systems with larger accretion rates, accretion heating within the disk overwhelms the heating from external starlight, so
that the waves no longer form or propagate.  The wave periods are comparable
to the heating and cooling time-scales, and are measured in years
 \citep{Watanabe2008}.

\begin{figure}[!ht]
\plottwo{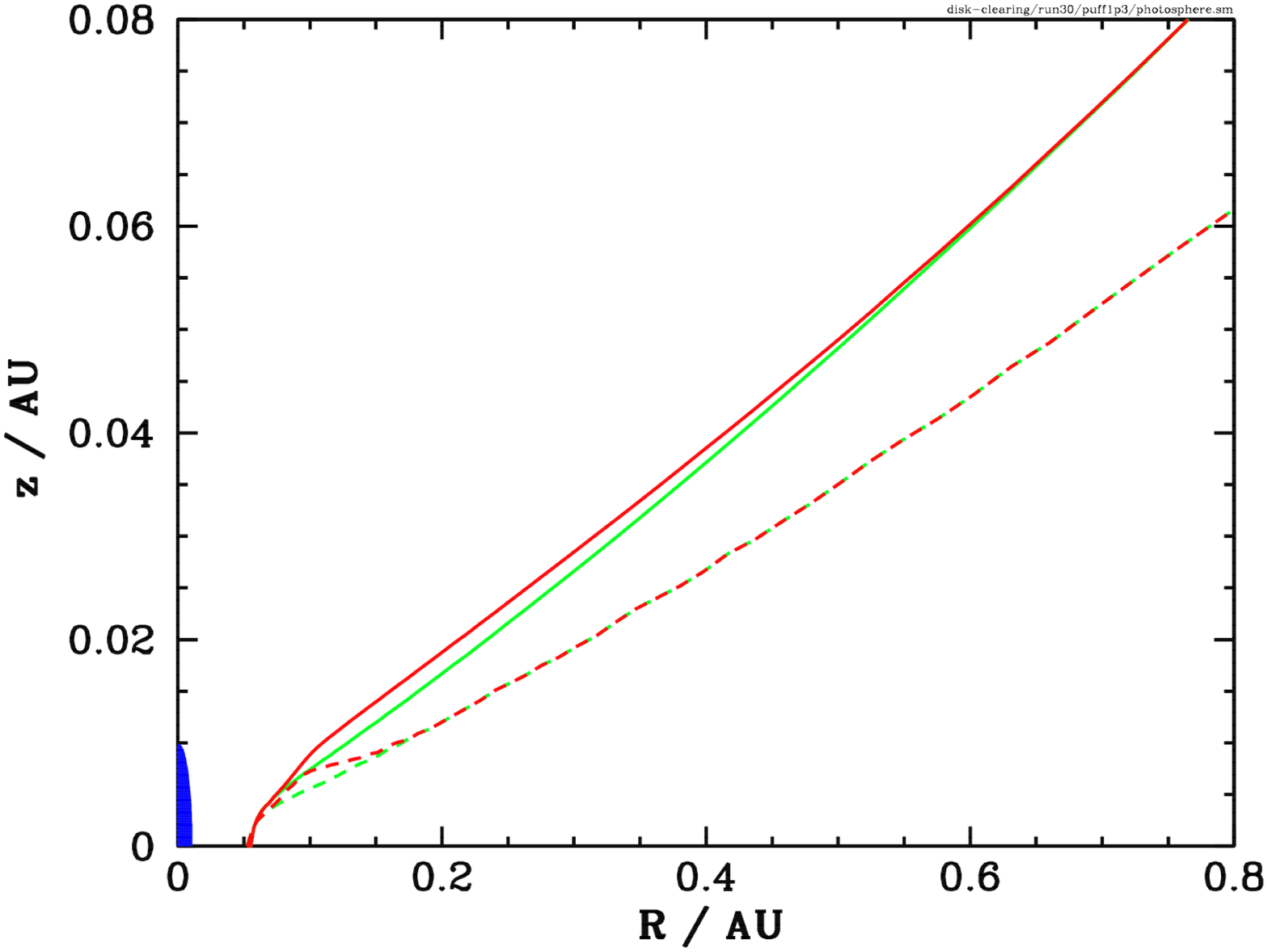}{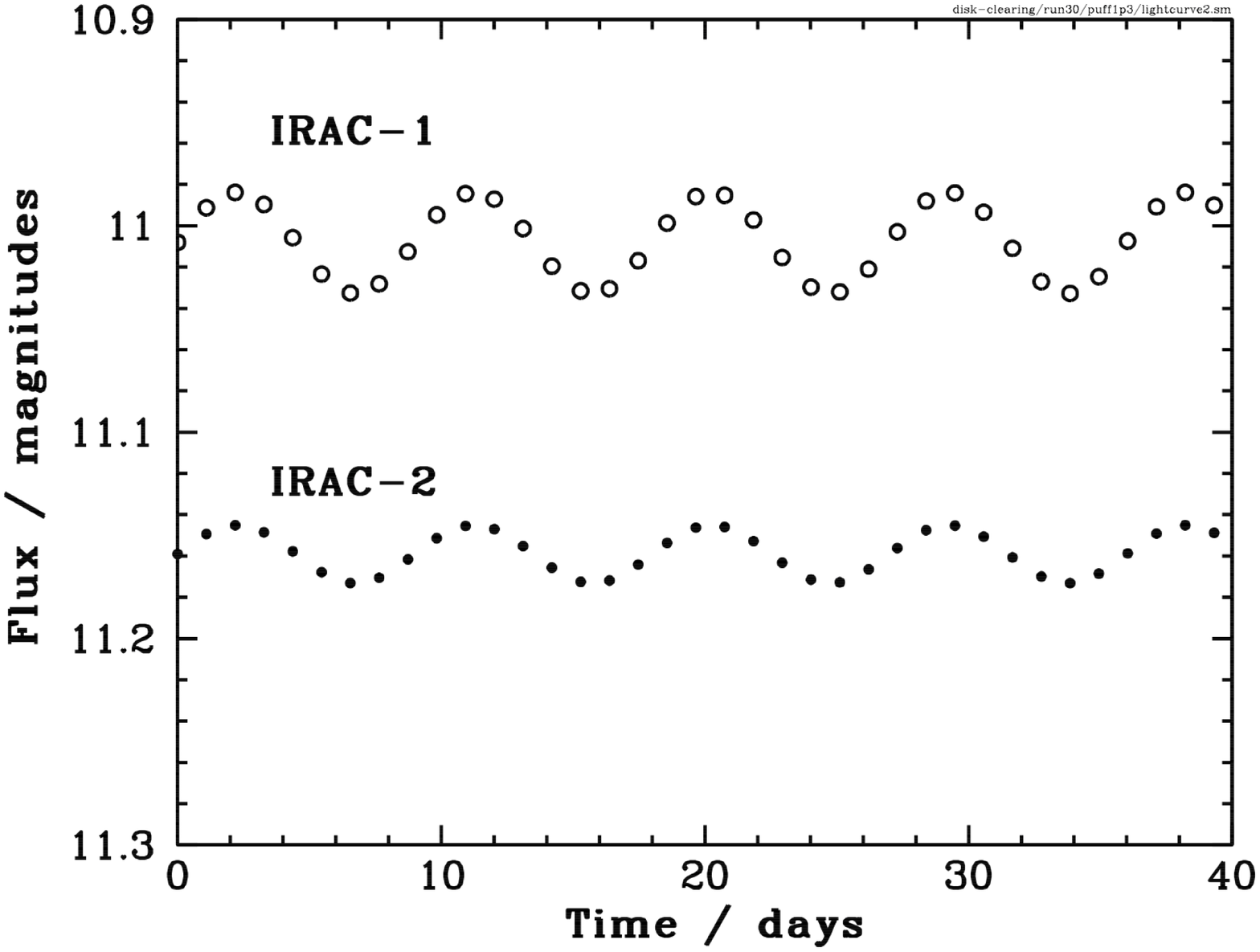}
\caption{{\it (Left) -- } A bump near the inner rim casts a long, cool shadow across the disk while itself being heated by the intercepted starlight.  Curves show the surfaces of unit optical depth for light coming from the star (solid) and from directly overhead (dashed), for cases with (red) and without a bump (green).  (Right) - Bumps whose height changes over time produce variability with amplitude greatest at shorter infrared wavelengths.  Synthetic IRAC 3.6 and 4.5 $\mu$m light curves for a model disk with a magnetically inflated inner rim.  (both panels, Turner et al., in prep). }
\label{eddies}
\end{figure}

But some T Tauri stars vary on time-scales of weeks, much shorter than
the thermal wave period.  The changes must arise near the disk inner
edge, where the orbital period is about a week.  On such short
time-scales, the outer disk can be treated as static.  A process
producing changes in the disk photosphere height over times comparable to
the orbital period is the same magneto-rotational turbulence that
drives the accretion flow onto the star.  The strength of the magnetic
fields varies episodically, leading to vertical excursions of the
disk photosphere by as much as 30\% \citep{Turner2010}.  The excursions are approximately
axisymmetric, at least over the length scales modeled in shearing-box
calculations.  When higher, the rim intercepts more starlight.  Monte
Carlo radiative transfer calculations \citetext{Turner N. J. et al., in
preparation} show raising the rim increases the area with
temperatures around 1000 K, while shadowing the disk beyond so that
the emission from cooler material is reduced.  The resulting
variations in the two warm-Spitzer bands are correlated, with
amplitudes generally decreasing toward longer wavelengths (see Figure \ref{eddies}).  Similar
effects can be expected from the shadow cast by the outer edge of the
gap opened in the disk by a giant planet, but the weaker frontal
illumination would tend to decrease the amplitude.

\section{Short wavelength Variability}

\subsection{Fabienne Bastien: High frequency variability of V1647 Orionis}

Numerous young stars have been observed to undergo large amplitude photometric outbursts.  While the observational record is still woefully incomplete, these events may play a significant role in the early evolution of many, possibly even most, young stars.  These outbursts have traditionally been divided into two major classes: FU Ori-like large amplitude ($>$ 4 mag), long duration (t$>$ 10s years) outbursts, with spectra dominated by absorption features characteristic of cool (GKM) supergiants \citep{Hartmann1996}, and EX Lup-like (often shortened to EXor) outbursts, which appear to be more frequent, of shorter duration, and characterized by heavily veiled spectra with strong emission and absorption components in accretion and outflow-sensitive features \citep{Herbig2007}.  

In 2004, the star V1647 Orionis was observed to undergo a major accretion outburst.  Numerous observations were obtained throughout the outburst, which persisted through 2006, but the source's photometric and spectroscopic properties resisted simple classification as a prototypical FU Ori-like or EX Lup-like outburst \citep[e.g., ][]{Fedele2007}.  Archival photometry revealed that the 2003 event was not V1647 Ori's first outburst, and additional objects have been observed to undergo outbursts with similar spectral characteristics \citep[e.g.,][]{Covey2010}, suggesting that these outbursts may signify a new phenomena of import for a star's early evolution.

Our study of high cadence time-series photometry of the 2003-2004 and
2008-2009 eruptions of V1647 Orionis revealed interesting
transient variability on time-scales of a few hours: we detect a highly
significant period of 0.13 days in our 2003 data that does not appear
in our 2009 data.  We attribute this period, which is inconsistent
with the object's rotation period, to a short-term radial pulsational
mode of the star excited by the sudden increase in its accretion rate
at the time of our 2003 observations.  This period, if also excited in
V1647 Ori's latest outburst event, would have presumably rung down by
the time we observed the star in 2009.  More near infrared and optical
observations with cadences of at least one image per hour over the
course of several days during the rise, plateau and decline phases of
such outbursts would allow us to better understand what short
timescale phenomena occur during these events.

\subsection{Ettore Flaccomio: Correlated optical and X-ray variability in accreting \\ Classical T Tauri Stars }

The magnetospheric accretion model, in which material is funneled from the inner edge of the circumstellar disk to the stellar photosphere along magnetic field lines \citep[e.g., ][]{Ostriker1995, Bouvier2007}, predicts that accreting material will be strongly shocked when it impacts the stellar surface.  This extremely hot \citep[T $\sim$ 10,000 K;][]{Herczeg2008} shocked region is believed to produce significant continuum emission at optical, ultraviolet, and X-ray wavelengths \citep{Gunther2007}.   Clear differences have indeed been detected between the X-ray properties of accreting Classic T Tauri stars (CTTs) and non-accreting WTTs: CTTs possess harder, more time-variable X-ray spectra than WTTs \citep{Flaccomio2006}, and detailed spectral studies suggest that accretion contributes a distinct, soft (2-3 MK) component to CTTs X-ray spectra \citep{Gudel2007}.  

Many open questions remain, however, as to the exact mechanisms underlying the connection between accretion and X-ray emission.  For instance, accreting systems are counter-intuitively seen to be significantly less X-ray luminous than predicted by standard magnetospheric accretion models \citep{JohnsKrull2007} as well as their WTTs counterparts \citep{Preibisch2005}, potentially due to the shocked emission being extincted by material in the overlying accretion columns \citep{Gregory2007}.  Previous observations have also failed to identify a clear correlation between CTTs optical and X-ray emission.  Accretion is expected to contribute a portion of the X-ray emission, originating in the photospheric accretion shock \citep[as opposed to coronal X-rays;][]{Stassun2006,Stassun2007,Grosso2007}.

In March 2008, \citet{Alencar2010} observed NGC 2264 with CoRoT for 23.5 days obtaining high-quality uninterrupted optical light-curves of its young stars. During the CoRoT pointing, two short Chandra observations were performed with a separation of 16 days, allowing us to study the correlation between optical and X-ray variability on this timescale, and thus the physical mechanism responsible for the variability \citep{Flaccomio2010}.  The variabilities of CTTs in the optical and soft X-ray (0.5-1.5 keV) bands are correlated, while no correlation is apparent in the hard (1.5-8.0 keV) band. Also, no correlation in either band is present for WTTs.  The correlation between soft X-ray and optical variability of CTTs can be naturally explained in terms of time-variable shading (absorption) from circumstellar material orbiting the star, in a scenario rather similar to the one invoked to explain the observed phenomenology in the CTT star AA Tau.  The slope of the observed correlation implies (in the hypothesis of homogeneous shading) a significant dust depletion in the circumstellar material.

\subsection{Marsden}

One of the key aspects of the solar dynamo is the reversal of the Sun's global magnetic field every $\sim$11 years. However, the dynamo mechanism for young rapidly-rotating solar-type stars is unclear and may well be fundamentally different to that of today's Sun \citep{Donati2003}.  To learn more about the type of dynamo operating in young stars we have used the technique of Zeeman Doppler Imaging \citep[ZDI;][]{SemelM:1989, DonatiJF:1997} to map the magnetic topology of two young solar-type stars (HD 141943 and HR 1817) over several years, searching in particular for evidence of polarity reversals in the global magnetic field. Given the rapid differential rotation seen on such stars \citep{MarsdenSC:2010a} it is expected that they have significantly shorter magnetic cycles than the Sun, perhaps as short as a few years.

\begin{figure}[!ht]
\plotfiddle{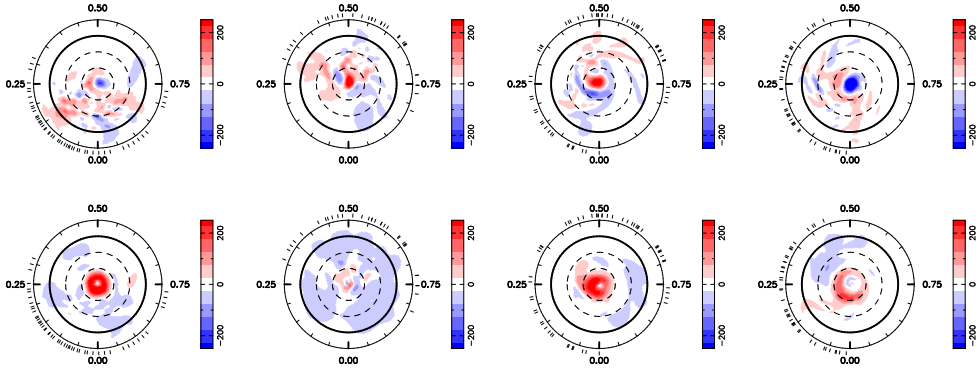}{2in}{-90}{50}{50}{-208}{150}
\caption{Preliminary radial (top) and azimuthal (bottom) magnetic field topologies for HR 1817 for the years (left to right), 2001, 2007, 2008 and 2009. The images are flattened polar projections extending down to -30$^{\circ}$ latitude. The bold line denotes the equator and the dashed lines are +30$^{\circ}$ and +60$^{\circ}$ latitude parallels. The radial ticks outside the plot indicate the phases at which the star was observed and the scale is in Gauss. (Marsden et al., in prep.)}
\label{b_fields}
\end{figure}

While HD 141943 shows evidence of changes in its magnetic topology it shows no evidence of a polarity reversal over the 3 years of observations \citep{MarsdenSC:2010a}. This result is similar to another young solar-type star (HD 171488) which has been previously observed using similar techniques \citep{MarsdenSC:2006, JeffersSV:2008, JeffersSV:2010}. In contrast, the preliminary results for HR 1817 indicate a reversal of the polarity of the star's radial magnetic field between 2008 and 2009, although a similar reversal is not seen in its azimuthal magnetic field (see Figure \ref{b_fields}; Marsden et al., in prep.). Such behavior suggests a phase delay in polarity reversal between the radial and azimuthal fields.

In order to assess how young solar-type stars undergo magnetic polarity reversals, optical spectropolarimetric observations of a statistically useful number of such stars needs to be undertaken regularly over a number years. A determination of how young solar-type stars go through magnetic polarity reversals and whether they have regular or chaotic magnetic cycles should greatly help our understanding of the dynamo in young Suns.

\section{Conclusions}

As this splinter summary demonstrates, recent years have seen considerable progress in our ability to observe, characterize, and model time-variable processes in the formation and early evolution of stars, circumstellar disks and planets.  These advances include the ability to study variability across a wider range of wavelengths and time-scales than previously possible, and to construct more detailed and computationally intensive models of star and disk processes.  

Nonetheless, our current understanding of time-variable phenomena in early stellar evolution is significantly incomplete.  In addition to the questions highlighted above, the audience identified several other outstanding challenges for studies of time-domain processes in the formation and early evolution of stars and planets.  These include:

\begin{itemize} 
\item{Improving the mechanisms and capabilities for conducting multi-site and multi-wavelength studies with long time baselines. }
\item{Achieving statistically robust constraints on variability (e.g., FU Ori outburst rate/duty cycle), rather than qualitative descriptions.}
\item{Identifying a critical mass of individual, rare systems (i.e, KH 15D, McNeil's Nebula, etc.) such that we can infer universal lessons from their specific properties.}
\item{Developing better observational and theoretical diagnostics to locate the source regions and determine the causes of variability.  Two examples are detecting kinematic signatures of the underlying gas motions in the line emission from the star and disk, and linking the photometric variability of the inner disk to changes in the scattered light from the spatially-resolved outer disk.}
\item{Understanding the implications for planet formation of time-variable structure in protoplanetary disks.} 
\end{itemize}

Meeting each of these challenges will require ingenuity, focused effort, and careful planning.  The young stars presenting here today suggest that we will indeed be up to the task, but only time\footnote{(bad) Pun intended.} will tell.

\acknowledgements 

We thank the Cool Stars 16 SOC for organizing a productive and enjoyable meeting.  We particularly thank the editors of this volume, whose enthusiasm and encouragement were critical to the ``timely'' preparation of this article.

\bibliography{Covey_K.splinter}

\end{document}